\documentclass[acmsmall,screen]{acmart}

\usepackage[whole]{bxcjkjatype}

\usepackage{afterpage}
\usepackage{arydshln}
\usepackage{array}
\usepackage{booktabs}
\usepackage{footmisc}
\usepackage{footnote}
\usepackage{lscape}
\usepackage{makecell}
\usepackage{multirow}
\usepackage{siunitx}
\usepackage{subcaption}
\usepackage{tablefootnote}
\usepackage{tabularx}
\usepackage{float}

\AtBeginDocument{%
  \providecommand\BibTeX{{%
    \normalfont B\kern-0.5em{\scshape i\kern-0.25em b}\kern-0.8em\TeX}}}

\setcopyright{none}

\acmConference{Under review}{February}{2024}

\begin{document}

\title[\proposed]{\proposed: Supporting Experts with a Multimodal Machine-Learning-Based Tool for Human Behavior Analysis of Conversational Videos}

\author{Riku Arakawa}
\authornote{All authors contributed equally to this research.}
\email{rarakawa@cs.cmu.edu}
\orcid{0000-0001-7868-4754}
\affiliation{%
  \institution{Carnegie Mellon University}
  \city{Pittsburgh}
  \state{Pennsylvania}
  \country{USA}
}

\author{Kiyosu Maeda}
\authornotemark[1]
\email{km9567@cs.princeton.edu}
\orcid{0000-0002-3270-1974}
\affiliation{%
  \institution{Princeton University}
  \city{Princeton}
  \state{New Jersey}
  \country{USA}
}
\affiliation{%
  \institution{ACES, Inc.}
  \city{Bunkyo}
  \state{Tokyo}
  \country{Japan}
}

\author{Hiromu Yakura}
\authornotemark[1]
\email{hiromu.yakura@aist.go.jp}
\orcid{0000-0002-2558-735X}
\affiliation{%
  \institution{University of Tsukuba}
  \city{Tsukuba}
  \country{Japan}
}

\renewcommand{\shortauthors}{Arakawa, Maeda, and Yakura}


\newcommand{\tabref}[1]{Table~\ref{#1}}
\newcommand{\figref}[1]{Figure~\ref{#1}}
\newcommand{\secref}[1]{Section~\ref{#1}}
\newcommand{\eqnref}[1]{Equation~\ref{#1}}
\newcommand{\appref}[1]{Appendix~\ref{#1}}

\newcommand{\eg}{\textit{e.g.},~}
\newcommand{\ie}{\textit{i.e.},~}
\newcommand{\etal}{\textit{et al.}~}
\newcommand{\etc}{\textit{etc}}
\newcommand{\proposed}{Providence}

\newcolumntype{C}[1]{>{\centering\arraybackslash}m{#1}}

\begin{abstract} 
Multimodal scene search of conversations is essential for unlocking valuable insights into social dynamics and enhancing our communication.
While experts in conversational analysis have their own knowledge and skills to find key scenes, a lack of comprehensive, user-friendly tools that streamline the processing of diverse multimodal queries impedes efficiency and objectivity.
To solve it, we developed \textit{\proposed{}}, a visual-programming-based tool based on design considerations derived from a formative study with experts.
It enables experts to combine various machine learning algorithms to capture human behavioral cues without writing code.
Our study showed its preferable usability and satisfactory output with less cognitive load imposed in accomplishing scene search tasks of conversations, verifying the importance of its customizability and transparency.
Furthermore, through the in-the-wild trial, we confirmed the objectivity and reusability of the tool transform experts' workflow, suggesting the advantage of expert-AI teaming in a highly human-contextual domain.
\end{abstract}



\begin{teaserfigure}
    \begin{center}
        \includegraphics[width=\textwidth]{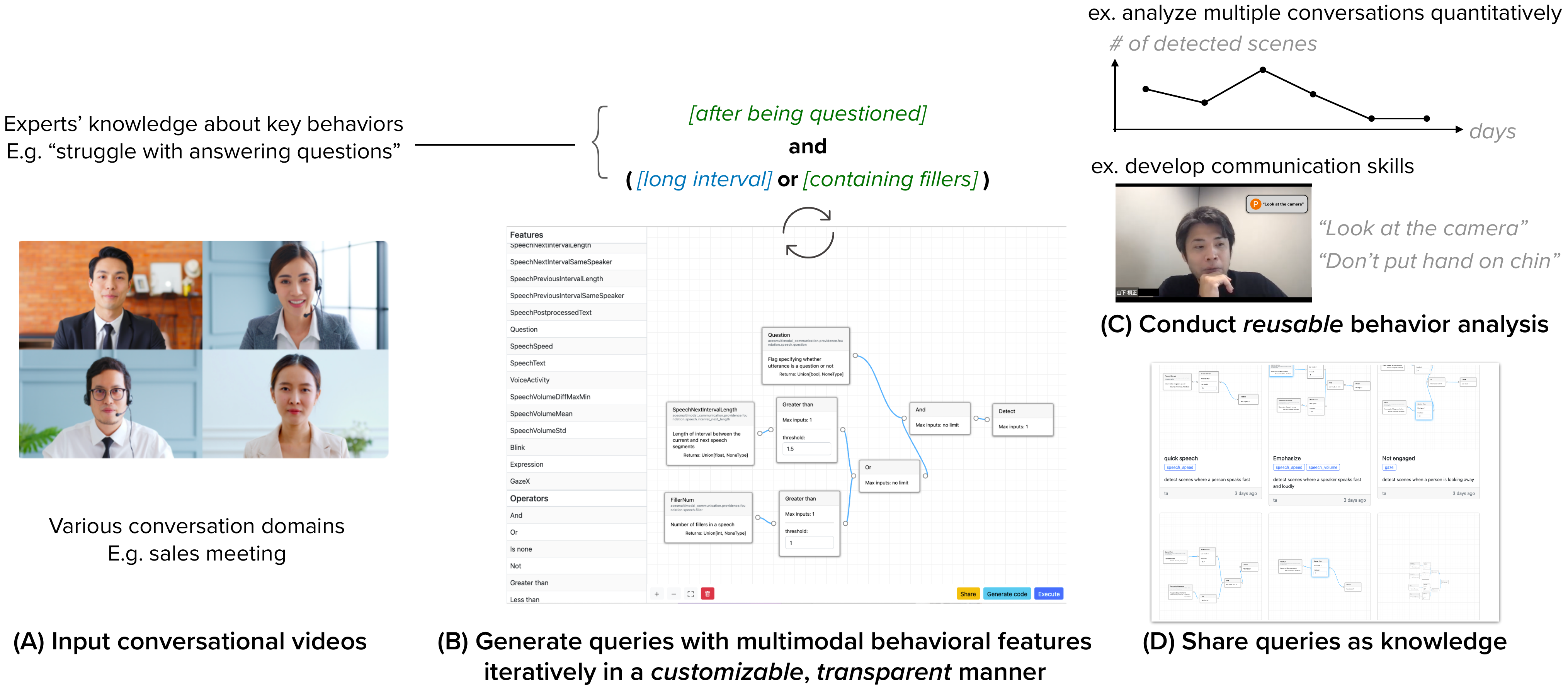}
    \end{center}
    \caption{Workflow of \proposed{}. (A) Experts in conversational analysis can input recorded videos in diverse conversation domains. (B) They create queries for various scene semantics by combining multimodal features via visual programming. For example, scenes where someone struggles with answering questions can be captured by combining linguistic and para-linguistic features. (C) Once queries are finalized, they can utilize them for their behavior analysis during conversations, such as monitoring target behaviors across multiple conversations. (D) Moreover, generated queries are sharable, enabling experts to collaboratively accumulate behavior analysis knowledge, which is often implicit and skill-dependent.}
    \label{fig:hero}
\end{teaserfigure}

\maketitle

\section{Introduction}
\label{sec:intro}

Human behavior and social dynamics during conversation have been investigated for a long time both in research and practice, resulting in myriad findings in social signal processing~\cite{DBLP:journals/taffco/VinciarelliPHPPDS12, Burgoon2017Social} and human activity analysis~\cite{DBLP:journals/csur/AggarwalR11}.
These insights have been applied to enhance communication and develop targeted interventions for individuals with specific needs~\cite{DBLP:conf/mfi/Gatica-Perez06}.
For instance, Patel~\etal\cite{DBLP:conf/chi/PatelHCPBR13} developed a system that captures nonverbal cues (\eg speech rate, facial expression) and provides real-time feedback to clinicians for enhancing empathic patient-centered care.
Arakawa and Yakura~\cite{DBLP:conf/chi/ArakawaY19} demonstrated that detecting anomalous scenes based on multimodal behavioral signals (\eg gaze) can assist professional coaches in executive coaching.
Recently, Deldjoo~\etal~\cite{DBLP:conf/sigir/DeldjooTZ21} introduced a generalized framework for retrieving scenes relevant to multimodal queries, serving as a foundational component for multimodal conversational information seeking.
As demonstrated, scene search of conversations based on participants' multimodal behavior holds significant potential for investigating social behavior and developing applications in diverse fields of dialogues.

However, a lack of tools comprehensively addressing the need for such multimodal searches poses challenges for many individuals attempting to conduct analyses.
While it is possible to write code to run machine-learning-based algorithms that capture behavioral signals (\eg gaze tracking) for each scenario, this approach excludes many experts (\eg clinicians, coaches) who can not necessarily code and hinders exploratory analyses.
Moreover, the incapability to holistically and iteratively test hypotheses on human behaviors raises validity concerns in the research findings~\cite{DBLP:journals/taffco/VinciarelliPHPPDS12} and leads to the gap between domain experts and software engineers when it comes to application development~\cite{DBLP:journals/corr/abs-2204-08471}.

In light of the supposition that the knowledge of HCI can be instrumental in addressing the gap, we first conducted a formative study involving semi-structured interviews with eight experts who regularly analyze human behavior in diverse dialogues.
While understanding the difficulty they face and confirming their need for computational support, we identified three design considerations: 1) \textit{customizability}, 2) \textit{transparency}, and 3) \textit{reusability}.
Customizability is important because the modalities and their relationship domain experts focus on change depending on their use case.
It is also essential to know which features and specific parameters affected the search results (\ie transparency) for them to interpret the results and modify queries exploratively.
Finally, the reusability of the search queries plays a fundamental role in collectively accumulating knowledge of human behavior analysis. 

Subsequently, we developed \textit{\proposed{}} as a prototype tool that meets the identified design considerations.
Its workflow is illustrated in \figref{fig:hero}.
At its core, \proposed{} embraces the visual programming paradigm~\cite{burnett1995visual} for constructing multimodal queries, wherein each behavioral feature is represented as a block and can be combined with various logic operators.
Its interface facilitates effortless customization across modalities, enables exploratory query modifications, and ensures transparent query representations.
Moreover, the generated queries can be exported and reused easily to develop standalone detection programs, expanding their applicability to various tasks beyond one-time searches, such as performing quantitative analyses across multiple conversations and developing real-time feedback systems.

We conducted a user study involving 12 participants to evaluate the effectiveness of \proposed{} in multimodal scene search of conversations recorded on a common video-conferencing platform.
Our study revealed that participants experienced reduced cognitive load in extracting key scenes from conversations relating to given descriptions of human behavior and expressed satisfaction with the scenes captured using the tool.
Additionally, a trend of the decreased time required to complete the search was indicated.
The results and the participant's comments corroborated the design of \proposed{}, bringing customizability and transparency in the process of multimodal scene search.

Moreover, we deployed \proposed{} in three in-the-wild situations of conversational analyses with 11 experts.
To facilitate sharing queries among experts, we introduced a \textit{knowledge-share repository} where users can contribute their generated queries and access those created collaboratively by others.
The study demonstrated that the tool started transforming experts' conversational analysis workflow by bringing objectivity and reusability to an open-ended, traditionally subjective domain.
These results underscore the potential of multimodal scene search of conversations and demonstrate the efficacy of \proposed{} in facilitating the process.

This paper makes several contributions:
\begin{enumerate}
    \item We identified the potential of deploying computational support in multimodal scene search of conversations through a formative study with experts in human behavior analysis.
    \item We developed \textit{\proposed{}}, a prototype that satisfies the identified design considerations by adopting the visual programming paradigm for combining various machine-learning-based algorithms to capture human behavior.
    \item We carried out a user study and demonstrated participants' preferences for the tool and reduced cognitive load in multiple multimodal scene search tasks, verifying the effectiveness of the tool's customizability and transparency.
    \item We deployed the tool in industrial organizations that regularly conduct conversational analysis, leading to experts' comments after the trial that shed light on how the objectivity and reusability offered by the tool affect their workflow.
\end{enumerate}

On top of the knowledge from HCI literature and through a dialogue with experts, we designed an intuitive, programmable analysis of human behavior during conversations.
This work demonstrates how the expert-AI teaming facilitated by the developed tool can open up their efficient and objective analysis workflow in the burgeoning domain that requires a significant degree of human context.

\section{Background and Related Work}
\label{sec:rw}

\subsection{Behavior Analysis during Conversation}
\label{sec:rw-behavior}

Beginning with Darwin~\cite{darwin1998expression}, many researchers have viewed human behaviors as spontaneous and unregulated expressions of internal states~\cite{DePaulo1992}.
As human conversation represents the most natural form of communication, extensive literature exists on behavior analysis during conversations in psychology and social science~\cite{sidnell2012handbook, Wooffitt2005}.
In HCI, researchers have utilized computers to quantify and process multimodal behavioral signals (\ie non-verbal ones such as facial expression and gaze, linguistic ones such as keywords and fillers, and para-linguistic cues such as speech tone and speed) to understand conversation dynamics~\cite{DBLP:journals/csur/AggarwalR11, DBLP:journals/taffco/VinciarelliPHPPDS12, DBLP:conf/hci/Otsuka11} and build systems~\cite{DBLP:journals/ijmms/NormanT91}.
Applications span various areas, including workplace meeting~\cite{DBLP:conf/icmi/NiheiNT17}, public speaking~\cite{DBLP:conf/iui/TanveerLH15, DBLP:conf/chi/DamianTBSLA15}, assistive technology~\cite{DBLP:conf/acii/HautBAZRNAH22}, health counseling~\cite{DBLP:conf/chi/PatelHCPBR13, DBLP:journals/tiis/RazaviSOAKH22}, education~\cite{DBLP:journals/imwut/AhujaKXVXZTHOA19}, executive coaching~\cite{DBLP:conf/chi/ArakawaY20, DBLP:conf/chi/ArakawaY19}, human assessment~\cite{DBLP:journals/corr/abs-2204-08471, Liebregts2019The}, and more (Refer to Table 1 in~\cite{DBLP:conf/icmi/Penzkofer0BMB21} for more examples).
These examples underscore the long-lasting interest in analyzing human behaviors during conversations and the potential for further advancements through computer technologies.

\subsection{Role of Scene Search of Conversations}
\label{sec:rw-role}

In behavior analysis during conversations, scene search or detection plays a pivotal role~\cite{DBLP:journals/mta/AlatanAW01, DBLP:conf/sigir/DeldjooTZ21}.
On the one hand, it is critical for developing interactive systems based on behavior analysis~\cite{DBLP:conf/iui/TanveerLH15, DBLP:conf/chi/DamianTBSLA15, DBLP:conf/uist/PavelGHA15, DBLP:conf/chi/ArakawaY19, DBLP:conf/uist/FraserMBDK20}.
For example, \textit{Rhema}~\cite{DBLP:conf/iui/TanveerLH15} is a public-speaking support system that provides real-time feedback based on the speaker's speaking patterns, such as when their speech volume or speed exceeds a pre-defined threshold.
On the other hand, scene search based on multimodal behavioral cues is also important in reviewing and analyzing conversations~\cite{DBLP:conf/icmi/TsfasmanFTLJO22, DBLP:conf/chi/ArakawaY20, DBLP:journals/corr/abs-2204-08471, DBLP:conf/chi/TakemaeOM04, DBLP:conf/sigir/DeldjooTZ21}.
Tsfasman~\etal~\cite{DBLP:conf/icmi/TsfasmanFTLJO22} discovered that non-verbal cues attendees remember can contribute to conversational memorability.
\textit{INWARD}~\cite{DBLP:conf/chi/ArakawaY20} demonstrated that anomaly scene detection during conversations could support executive coaches reviewing sessions by offering semantically meaningful scenes.
Similarly, AI-driven scene extraction can assist professional assessors in making complex decisions (\eg suitability for a specific job) about the individuals they interacted with~\cite{DBLP:journals/corr/abs-2204-08471}.

Despite the widespread adoption of scene search of conversations for both practice and research, no standardized tool is available, making the analysis process time-consuming and challenging for experts.
Moreover, the lack of common methods hinders the easy transfer of findings; such specificity is known as one of the open issues in this research field~\cite{Burgoon2017Social}.
Given that, we presumed that a tool that comprehensively addresses scene search of conversation based on multimodal behavioral cues is in high demand, which motivated us to conduct the formative study.
We also expected that such a tool could also become a common analysis method for related disciplines (\ie social psychology), similar to how \textit{DeepLabCut}~\cite{Mathis2018DeepLabCut} has become indispensable in neuroscience.

\subsection{Tools for Conversational Analysis}

On the other hand, prior HCI work has proposed approaches for visualizing multimodal features in specific conversational situations~\cite{DBLP:conf/huc/HoqueCMMP13, DBLP:journals/imwut/SamroseZWLNLAH17, DBLP:journals/ijmms/Ez-zaouiaTL20, DBLP:conf/chi/SamroseMSSRHRMC21, DBLP:conf/uist/LiCTC21, DBLP:conf/icmi/Penzkofer0BMB21,  DBLP:journals/pacmhci/BenkeSLM22}, rather than equipping search functions.
These approaches help users review or analyze conversations by examining relevant feature information that is automatically extracted.
For example, \textit{MeetingCoach}~\cite{DBLP:conf/chi/SamroseMSSRHRMC21} is an intelligent dashboard providing transcripts and multimodal cues of attendees, such as when their speaking tone changes in meetings.
\textit{MACH}~\cite{DBLP:conf/huc/HoqueCMMP13} is an interview-practice system that presents feedback on multimodal aspects during conversations, such as speech tone and pauses in time series.
Li~\etal\cite{DBLP:conf/uist/LiCTC21} achieved hierarchical summarization of video contents based on the language information and presented it to users for an efficient review of talk videos.
\textit{Emodash}~\cite{DBLP:journals/ijmms/Ez-zaouiaTL20} is an interactive dashboard that displays learners' emotional state to support tutors in video-based teaching.

However, these visualization-based approaches were developed for specific purposes (\eg providing feedback on interview practice), and the features to display were often determined arbitrarily.
As a result, these solutions lack flexibility, as users cannot customize the features for their own needs.
Additionally, these tools are intended to support one-shot scene extraction, that is, finding a specific scene from a visualization (\eg looking for moments when speech speed is high in its time-series graph).
Thus, they do not apply to performing scene extraction across various situations with consistent logic.
In other words, this introduces subjectivity into the process, which is often considered a limitation in the traditional analysis by experts~\cite{DBLP:journals/corr/abs-2204-08471}, and demands a tool enabling customizable scene search to accommodate various applications in conversational analysis.

\subsection{Techniques for Scene Search of Video}

Many studies have addressed video retrieval, such as on TV programs, movies, and security cameras~\cite{DBLP:journals/corr/abs-1205-1641, DBLP:journals/corr/ZhouLT17, DBLP:conf/mir/2018lsc}.
For dialogue videos, a straight-forward approach is text-based, where users can search scenes by keyword, as in~\cite{DBLP:conf/uist/PavelGHA15, DBLP:conf/chi/ZhuWYSZHZMS17}.
At the same time, multimodal scene search has been addressed through semantic alignment between queries and scenes~\cite{DBLP:conf/mmm/MoumtzidouMAMIG16, DBLP:conf/mmsp/RotmanPA17}.
For example, \textit{VERGE}~\cite{DBLP:conf/mmm/MoumtzidouMAMIG16} achieved a search of videos by using visual similarity.
Similar methods have been proposed to deal with the semantic search of lifelogging videos~\cite{Aizawa2004Efficient, DBLP:conf/mir/MejzlikVKSL20, DBLP:conf/mir/AlamGG21, DBLP:conf/mir/SpiessS22}, where queries often contain names of objects or locations (\eg ``red car''~\cite{DBLP:conf/mir/AlamGG21}).

Such semantic-based approaches, however, cannot be directly applied to the scene search of conversations based on behavioral cues.
While we will have a more detailed discussion based on our formative study, we can raise two reasons here.
First, behavioral cues are often not as striking as conventional scene descriptions (\eg using object names).
Secondly, end-to-end approaches lack interpretability of the extracted scenes, especially in conversational analysis, where queries can be more ambiguous than conventional video retrieval.
For example, it is common to try to identify scenes where a person struggles with question-answering~\cite{degand2019should}.
In such cases, it is difficult even for humans to agree on the concrete definition of the optimal query.
Thus, it is essential to inform users of the reasons behind the outputs and allow them to explore the data to maintain their trust in the system~\cite{DBLP:conf/chi/AmershiWVFNCSIB19}.

\section{Formative Study}
\label{sec:formative}

From the above literature, we figured out not only the potential of introducing the knowledge of HCI to this burgeoning domain of behavior analysis during conversations but also the expected shortcomings in merely applying the existing techniques.
This prompted us to have a deeper understanding of the analysis process in practice and the difficulty being faced, and thus, we conducted semi-structured interviews with domain experts.
We also tried to derive key design considerations to develop an appropriate system in this emerging domain.

\subsection{Semi-Structured Interviews}

We used personal contacts and word of mouth to recruit experts who regularly check or analyze human behavior during conversations in different professions and who did not know about our project.
In the end, eight people agreed to join our phone-based interviews (See \tabref{tbl:interviewees} for their expertise).
Here, we have made the utmost effort to gather experts who are likely conducting analyses in various professions.
Still, we cannot deny the possibility of disciplines not included here; \eg interrogation investigators could apply a similar practice.
On the other hand, we found that the interviews revealed findings that could be extrapolated to other contexts, and thus, we moved forward to develop a computational support tool based on the results.
In this context, we should mention that assessing the scope of application for the findings from this formative study is a part of important future work, as mentioned in \secref{sec:discussion-potential}.

\begin{table}[t]
  \caption{Backgrounds of the participants in the semi-structured interviews.}
  \label{tbl:interviewees}
  \begin{tabular}{lllll}
    \toprule
    ID & Gender & Role & Domain & Job experience \\
    \midrule
    P1  & M & assessor & human assessment & 5 years \\
    P2  & F & coach & executive coaching & 3 years \\
    P3  & F & counselor & mental health counseling & 2 years \\
    P4  & M & salesperson & sales meeting & 1 years \\
    P5  & M & interviewer & job interview & 7 years \\
    P6  & M & team manager & internal meeting & 5 years  \\
    P7  & F & researcher & social psychology & 4 years \\
    P8  & M & researcher & social signal processing & 4 years \\
    \bottomrule
  \end{tabular}
\end{table}
The interviewees were asked the following questions: ``\textit{What is the purpose of your reviewing or analyzing conversation with a focus on human behavior?}''; ``\textit{How do you review or analyze it?}''; ``\textit{What are the characteristics of scenes you want to focus on in the reviewing process?}''; ``\textit{Is there any problem in the reviewing process?}''; ``\textit{What kind of support would you need from computers?}''.
Each interview took approximately 20 minutes.

\subsection{Results}
\label{sec:formative-results}

\subsubsection{Purpose of reviewing conversation}

The interviewees mentioned various purposes in response to the first question, ``\textit{What is the purpose of your reviewing or analyzing conversation with a focus on human behavior?}''
There are two overarching common directions at a high level.
P1, P2, P3, P7, and P8 watch conversational videos to facilitate the process of structuring objective thoughts (\eg testing their hypotheses or evaluations) regarding the attendees in the conversation.  
\begin{quote}
    I try to verify the hypotheses that come from my subjective impression of the person (in human assessment). [P1]
\end{quote}
\begin{quote}
    We analyze clients' behavior during the coaching session from an objective perspective. This is done to ensure that no potentially overlooked scenes are missed. [P2]
\end{quote}
\begin{quote}
    I check the video to see if there are any missing signals my client possesses. [P3]
\end{quote}
\begin{quote}
    I try to find specific patterns in people's behavior by watching the video carefully. [P7]
\end{quote}
\begin{quote}
    I check multiple videos to find a pattern before specific behavior emerges. I then try to make a system to detect such patterns. [P8]
\end{quote}

On the other hand, P4, P5, and P6 review videos to gather insights for improving their own and others' behaviors during conversations.
\begin{quote}
    We review each other's online sales videos. We learn from senior members and provide feedback to subordinates. [P4]
\end{quote}
\begin{quote}
    We review job interviews conducted by other members, especially by junior members, to try to give them feedback. [P5]
\end{quote}
\begin{quote}
    My team sometimes checks the engagement of each attendee in the recorded meeting videos to enhance the approach to internal meetings. [P6] 
\end{quote}
While the interviewees' areas of expertise are diverse, a commonality in reviewing conversational videos is observing them based on specific criteria concerning participants' behaviors.
Conversely, their purpose is distinct from remembering their past conversations or catching up on unattended conversations, for which they can mostly rely on the (automated) transcriptions.
Rather, the responses revealed a separate motivation for reviewing conversational videos from an objective viewpoint.

\subsubsection{How they review conversation}

The responses to the next question, ``\textit{How do you review or analyze conversation?}'', further provide insights into their practice of reviewing the conversation. 
Specifically, they look for scenes during conversation based on behavioral cues of specific people.
\begin{quote}
    I usually inspect the assessee's verbal and non-verbal behaviors and try to find cues for making the final evaluation of them. This is because, when I conducted interviews with an assessee, it was difficult to remember all cues to make the decision at the moment. I often take rough notes very quickly and review the scene afterward for reflection. [P1] 
\end{quote}
\begin{quote}
    When checking sales meetings conducted by my subordinates, I find out what questions the customer asked and what points they were engaged in most. [P4]
\end{quote}
\begin{quote}
     For instance, the juniors often look angry and speak fast in the interviews. I often look for such scenes [to provide feedback because it is negative behavior]. [P5]
\end{quote}
As shown in the responses, the scenes they look for involve diverse cues.
This point was further stressed in response to the next question, ``\textit{What are the characteristics of scenes you want to focus on in the reviewing process?}''.
Most of them regarded multimodal behavioral cues as an important aspect;
\begin{quote}
    There is sometimes a contradiction between what a coachee says and what their non-verbal behavior represents. Those scenes are often good moments to dig deeper as the coachee may want to expose more of themselves. Thus, I monitor their behaviors such as gaze and postures as well as speech patterns such as tone in the conversation. [P2]
\end{quote}
\begin{quote}
    I pay attention to scenes when their non-verbal and para-linguistic behaviors change as they contain important information to assess their mental state. Specific characteristics vary depending on the case, but often facial expression tells a lot. [P3]
\end{quote}
\begin{quote}
    I analyze human behavior multimodally. The relationship between internal states and behavior is complex and I often playback scenes exploratively. [P7]
\end{quote}

The fact that multimodal behavioral signals are utilized for various purposes aligns with the prior work discussed in \secref{sec:rw-behavior}.
Consequently, we hypothesized 1) that a tool for scene search based on behavioral cues would help the analysis, and 2) that such a tool without sufficient \textit{customizability} would not adequately address their needs.

\subsubsection{What support they want}

Then, we observed several shared needs in their responses to the questions, ``\textit{Is there any problem in the reviewing process?}'' and ``\textit{What kind of support would you need from computers?}''.
First, they expressed a desire for computational support to make the process of reviewing more efficient; while they usually watch entire conversations manually, such a way takes a significant amount of time.
Moreover, they mentioned a concern about subjectivity in the process due to the lack of quantified metrics; for instance, 
\begin{quote}
    I am often told by my manager that I tend to get stuck and look restless in answering customers' questions. But, by just watching the recorded videos myself, I am not sure when I do it and whether I'm actually making an improvement to stop it over time. [P4]
\end{quote}
\begin{quote}
    I feel each member relies on their own criteria in finding key scenes to make a summary, which often leads to highly subjective conclusions. [P6]
\end{quote}

Subsequently, when the interviewer mentioned the possibility that computers automatically extract such scenes for them, all interviewees reacted positively to the aim of our project, heaping their ideas on it; for example,
\begin{quote}
    That is cool. While the members do have their knowledge to analyze behavior, they cannot do programming. A good combination is a promising direction. [P6]
\end{quote}
\begin{quote}
    Such a tool will be a great help. I want it to tell me why each scene is extracted as well. We often look at different characteristics of behaviors, so it is important to see the clear reason in the results and make the iteration process more efficient. [P8]
\end{quote}
These comments suggested facilitating domain experts' analysis with a computational tool would be beneficial.
Moreover, such a tool is desired to offer \textit{transparency} in the scene search results.
This point echoed with prior findings on human-AI interaction in the sense that we need to enable users to interpret the results from computers~\cite{DBLP:conf/chi/AmershiWVFNCSIB19}, especially in situations with high human context, like conversation~\cite{DBLP:journals/corr/abs-2204-08471}.

Moreover, P4 and P5 found another benefit of making the analysis process computational.
\begin{quote}
    It would be nice if we could search scenes across different meetings with the same criteria. That would allow me to monitor my behavior quantitatively, and the results would be more trustworthy. [P4]
\end{quote}
\begin{quote}
    I would love it if it allowed us to compare different candidates with the same criteria, removing the subjectivity and skills of each interviewer. [P5]
\end{quote}
These comments highlighted the importance of ensuring that the analysis is \textit{reusable} across multiple conversations to maintain consistency in the results.

\subsection{Conclusion}
\label{sec:formative-conclusion}

The interviewees in this study, coming from diverse backgrounds, were involved in reviewing conversational videos as part of their work.
Their objectives varied from testing hypotheses about participants to improving participants' behaviors. 
However, a common process identified across these activities is looking for scenes with specific criteria regarding participant behaviors in the videos.
Furthermore, they share challenges in the current process, such as its highly subjective and time-consuming nature.
Therefore, we concluded that there is a solid need for a computational support tool to facilitate this process, targeting those who engage in these activities with domain-specific knowledge.
Specifically, the benefit of such a tool includes objectivity and easiness of iterative analysis regarding human behavior.
We also found that the cues interviewees look at are mainly individual behavior, for instance, how a client behaves during a coaching session.
This point guided us to develop a tool that computationally associates each attendee's behaviors with scenes in conversational videos.
Moreover, the tool should be able to simultaneously treat various behavioral features, such as linguistic, para-linguistic, and non-verbal, as well as support a nonlinear combination of those cues (\ie the logical relationships between the cues not necessarily be linear but can be a tree-like structure).

Based on the interviewees' comments, we derived the following design considerations over three facets:
\begin{enumerate}
    \item \textbf{Customizability}: Given there are various use cases using a diverse set of multimodal behavioral signals (linguistic, para-linguistic, non-verbal), the tool must allow users to customize their queries easily. Specifically, it needs to consider the nonlinearity of the combinations of the behavioral signals, as their number or order could not be predetermined in the exploratory processes.
    \item \textbf{Transparency}: The tool must offer a means for users to interpret the results to maintain users' trust in the tool during the complex processes based on their implicit knowledge. Specifically, it needs to present the logic behind the scene extraction in an inspectable manner, such as each behavior feature contributing to the result.
    \item \textbf{Reusability}: The search queries should be applicable to other conversations to make the analysis consistent and comparable. Specifically, it is desirable to maintain interoperability, as the queries would be applied to videos to be recorded in the future.
\end{enumerate}
The factors share the concept discussed in the recent studies on human-AI interaction~\cite{DBLP:conf/chi/AmershiWVFNCSIB19}.
While the mere terms of the above considerations seem not novel, we believe in the importance of connecting the vocabulary in HCI research with experts' voices in a burgeoning domain, especially in highly contextual situations like human behavior analysis where technology support has not been well-explored.
Based on the design considerations, we propose an artifact and discuss how each element contributes to solving the identified issue in the rest of the paper.

\section{\proposed{}}

Based on the conclusion, we designed \textit{\proposed{}} as a tool for multimodal scene search of conversations.
One of the remarkable points in its design is the adoption of visual programming.
In this section, we present the rationale behind the adoption and the detailed explanation of the interface and implementation of \proposed{}.

\begin{figure*}[t]
    \begin{center}
        \includegraphics[width=\textwidth]{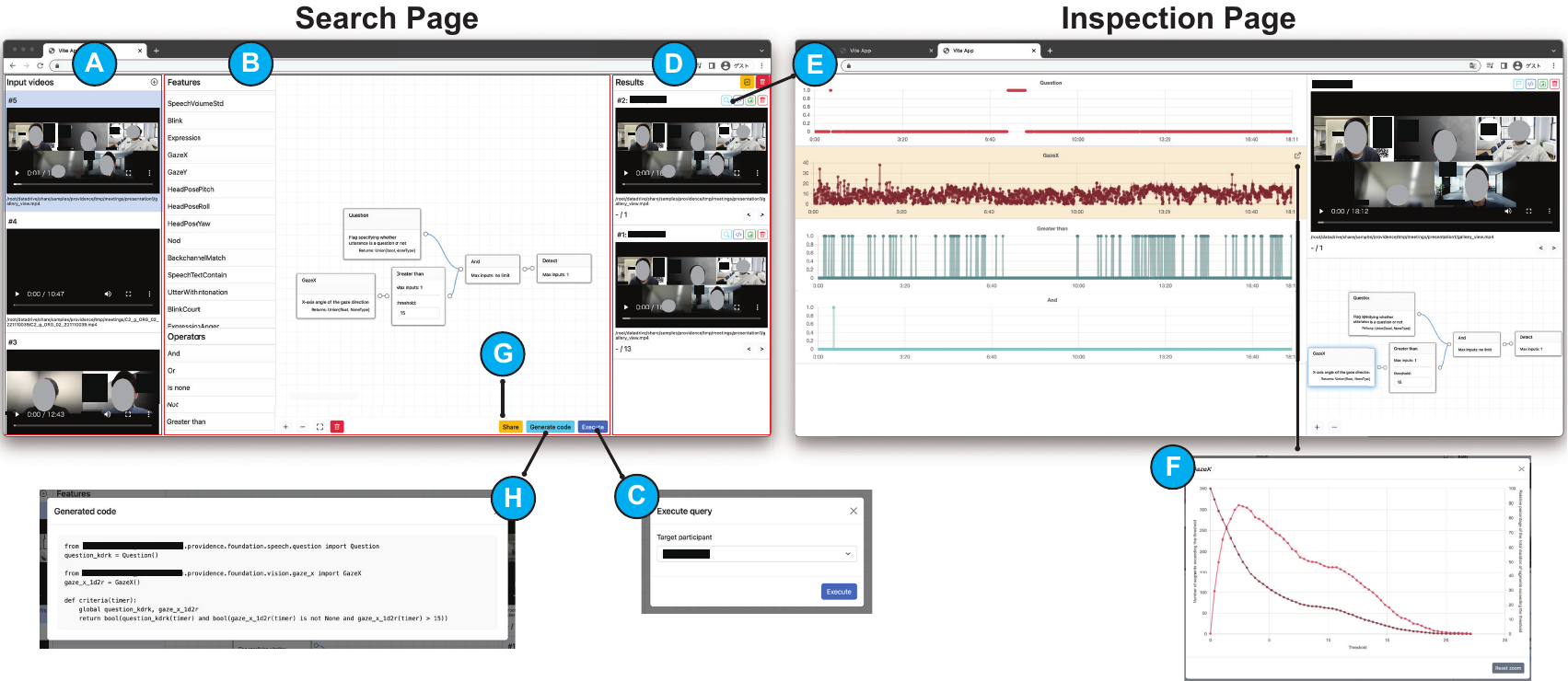}
    \end{center}
    \caption{Interface of \proposed{} with an example query that detects scenes when a person is questioning but not looking straight at the screen according to one's gaze. Some parts (\eg attendees' faces) are anonymized for blind review.}
    \label{fig:interface}
\end{figure*}

\subsection{Why visual programming?}

The potential of visual programming~\cite{burnett1995visual} has gained attention in recent studies on machine-learning prototyping, as it offers users a transparent visualization of the logical relationship between the features and allows them to iteratively customize algorithms~\cite{DBLP:conf/chi/CarneyWAPHGJPC20}.
Visual programming can be used easily and intuitively by those who are not programming experts to design and apply machine learning algorithms, compared to other tools that force users to write codes, such as~\cite{DBLP:journals/corr/abs-1910-02993}.
This is especially beneficial when connected with domain experts, such as in biomedical image analysis~\cite{godec2019democratized}, prompt engineering for large language models~\cite{DBLP:conf/chi/WuTC22}, and multimedia authoring~\cite{RapsaiDu}.

These studies suggest that visual programming can enhance the transparency and controllability of complex machine-learning algorithms in different domains, motivating us to its adoption in human behavior analysis during conversation.
Specifically, as discussed in \secref{sec:formative-conclusion}, our situation requires a high level of customizability, as it involves the combination of nonlinear multimodal behavior signals calculated through various machine-learning algorithms.
In other words, we expected that other approaches (\eg parametric design~\cite{Koyama2018,DBLP:conf/ijcai/Yakura0G21}) could be insufficient to perform such a complex process.
Furthermore, visual programming can offer transparency in the analysis process by visualizing the logical relationship of behavior signals.
This is highly beneficial for determining thresholds or relevant parameters across multiple signals, which is necessary for finding optimal search queries.

Importantly, visual programming is crucial also in terms of maintaining reusability.
Here, \proposed{} associates such visual representation with an equivalent Python code, as described later.
This thus ensures the reusability of artifacts made by experts, even for videos to be recorded in the future and even without the interface by running the code in a standalone manner, which promotes a collective accumulation and exploitation of their knowledge.

\subsection{Interface}

To fully leverage the advantages, we designed the interface of \proposed{} as shown in \figref{fig:interface}.
Since the details of the features are described in \secref{sec:proposed-impl}, we here provide a quick walk-through.
(A) Users can use various conversational videos as input for the tool.
(B) They can select features from a diverse set and customize the relationship of the selected features (\eg threshold, filtering logic) in a visual block representation to form a query.
(C) They can execute a query by choosing an attendee in the conversation as an analysis target.
(D) In the right column, the extracted scenes according to the query are presented, and users can play back each scene with a corresponding transcription.
(E) When they click the inspection button near the result, they can see the time-series visualization of each feature used in the query.
Here, they can see not only the values of raw features but also when scenes are detected for each modality, \eg based on the given threshold.
(F) Furthermore, they can see how each parameter in the query affects the result so that they can find the optimal value instantly.
(G) Once they make a query, they can export it in a JSON representation to share the knowledge with others.
(H) They can also export it in Python code to deploy it as an instant application.

\subsection{Implementation}
\label{sec:proposed-impl}

\begin{figure*}[t]
    \begin{center}
        \includegraphics[width=0.8\textwidth]{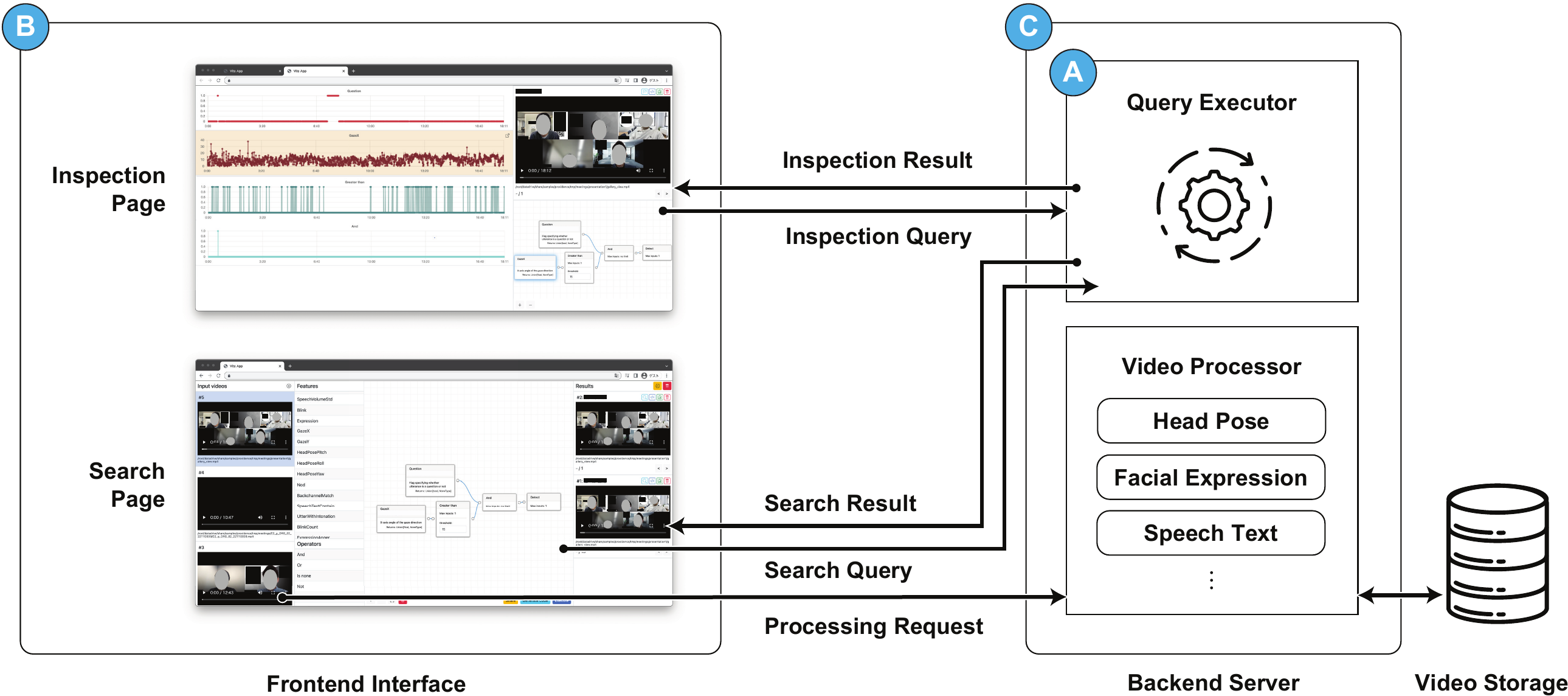}
    \end{center}
    \caption{Architecture of \proposed{}, which consists of (A) a programmatic framework for querying conversational videos, (B) a frontend interface with visual programming and feature visualization, and (C) a backend server for machine-learning algorithms and query processing.}
    \label{fig:architecture}
\end{figure*}

We implemented \proposed{} as a web-based tool, as illustrated in \figref{fig:architecture}, consisting of (A) a Python-based programmatic framework for querying conversational videos, (B) a frontend interface with visual programming functionality in Vue.js and (C) a backend server in Python.

\subsubsection{Framework for Querying Conversational Videos}

\begin{figure}[t]
    \begin{center}
        \includegraphics[width=0.5\linewidth]{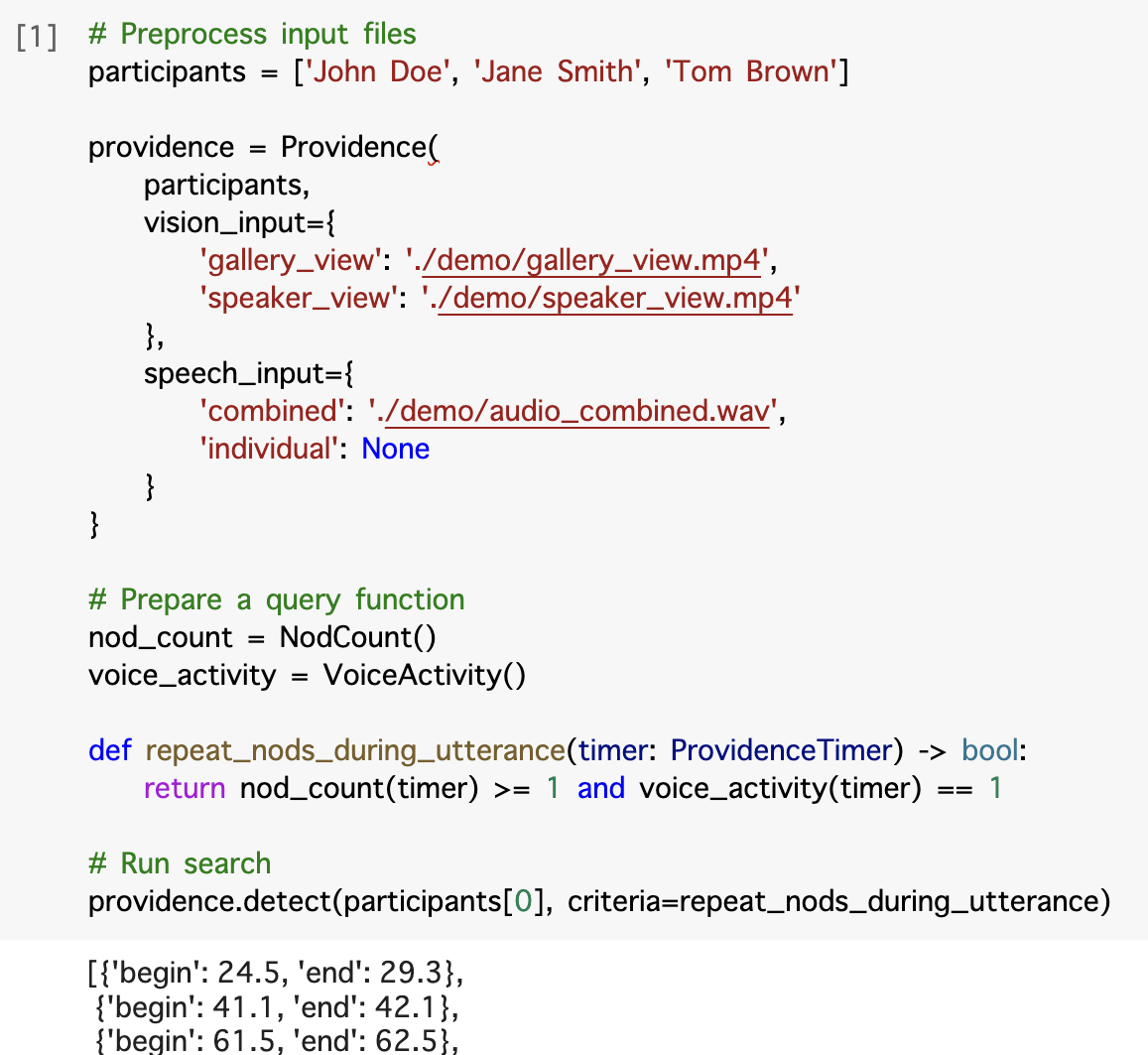}
    \end{center}
    \caption{An example code of searching scenes with \proposed{}'s programmatic framework. In this example, two features (\ie nod count and voice activity) are combined. The input format was enabled to be flexible to accept videos recorded on common video-conferencing platforms while maintaining a consistent output format.}
    \label{fig:code}
\end{figure}
\begin{table}[t]
\caption{List of multimodal features available in \proposed{}. Detailed implementation is provided in the Appendix.}
\label{tbl:features}
\begin{tabular}{lll}
\toprule
Non-verbal                                       & Linguistic                                                 & Para-linguistic           \\
\midrule
Facial keypoints \cite{DBLP:conf/iccv/FangXTL17}                & Speech text \cite{DBLP:conf/interspeech/GulatiQCPZYHWZW20} & Voice activity            \\
Body pose \cite{alphapose}                       & Filler                                                     & Speech frequency \\
Head pose \cite{DBLP:conf/cvpr/DengGVKZ20}       & Backchannel                                                & Speech volume    \\
Gaze direction \cite{DBLP:conf/eccv/FischerCD18} &                                                            & Speech length             \\
Facial expression \cite{DBLP:conf/cvpr/HeZRS16}  &                                                            & Speech speed              \\
                                                 &                                                            & Speech interval            \\[0.2em]
\cdashline{1-3} & & \vspace{-0.8em}\\
Nod                                              & Question                                                   & Intonation          \\
Blink                                            & Sentiment                                                  & Speech overlap            \\
\bottomrule
\end{tabular}
\end{table}

To guarantee customizable and reusable searches, we prepared a programmatic framework that unifies various features extracted from conversational videos.
\figref{fig:code} shows an example code of searching scenes when a person nods while speaking.
In this framework, search queries are represented as a function that takes a \textit{timer} object as its input and returns a flag indicating whether the specific timing satisfies the search criterion.
As this example code implies, our framework performs searches based on time-based sampling.
Indeed, it is sensible to perform an utterance-based search to capture utterances that satisfy a given query, \eg questioning utterances.
However, we need to deal with various features, some of which would be too coarse at the utterance level to capture, such as one's facial expression.
Thus, we implemented \proposed{}'s framework to align all features based on the timestamp of input videos and capture time segments that satisfy a given query.
For example, when a user builds a query to specify scenes where the number of fillers exceeds $N$, all segments of utterances that contain $N$ or more fillers will be captured.
When the user also applies a filter on the blink feature in the query, scenes where a person is blinking will be captured among the segments.

Also, our engineering efforts made it possible for the module to be flexible with the input video formats such that most video-conferencing platforms are supported (\eg \textit{Zoom}, \textit{Google Meet}, and \textit{Microsoft Teams}).
Namely, it can accept either (or both) gallery-view or speaker-view video as vision input and either (or both) combined or individual audio input, as shown in \figref{fig:code}, outputting features in a consistent format.
Thus, theoretically, our framework has no limit regarding the number of people in conversation, as long as the video-conferencing platforms can serve the number of people.
We also note that it identifies the correspondence between each person in the video and the provided list of the names of participants by recognizing the name text that appears in the video.
Therefore, when some people join a video-conferencing platform from the same room, our framework differentiates them by assigning pseudo names (\eg \textit{Speaker A}) based on speaker diarization \cite{DBLP:conf/icassp/BredinYCGKLFTBG20} and face tracking \cite{DBLP:conf/iccv/FangXTL17}.
In addition, when some people turn off their videos, it extracts only linguistic and para-linguistic features for them.

\tabref{tbl:features} shows a list of features we implemented in \proposed{}, which are calculated through the video processor module.
We chose those features based on existing work (\eg \cite{DBLP:conf/huc/HoqueCMMP13,DBLP:conf/chi/SamroseMSSRHRMC21}) and the responses from the participants in the formative study.
Note that we can easily add a new feature to the video processor thanks to its modular architecture.
The video processor module first applies a speech recognition pipeline (including speaker diarization) and para-linguistic computation.
It also applies vision algorithms (\eg head pose estimation through facial keypoints detection) sequentially to all of the frames in a video.
After that, it gathers all features calculated from the outputs of the pipeline.
For example, we trained a Transformer model~\cite{DBLP:conf/nips/VaswaniSPUJGKP17} to detect questioning utterances and sentiment from the transcriptions.
Each model was mostly adopted from existing machine-learning algorithms, and we describe the implementation details, accuracy information, and processing speed of each model in the Appendix (\secref{app:implementation}).

\subsubsection{Frontend Interface}
\label{sec:proposed-impl-frontend}

On top of the programmatic framework, we built a frontend web interface for intuitive search by introducing the visual programming paradigm.
As shown in \figref{fig:interface}B, search queries are represented as a diagram consisting of several blocks, each of which corresponds to a single feature or a logic operator.
Users can edit queries, for example, by applying a specific threshold or rule to features (\eg \texttt{Greater} \texttt{than}, \texttt{Less} \texttt{than}, and \texttt{Text} \texttt{contain}) or combining multiple features (\eg \texttt{And}, \texttt{Or}, and \texttt{Not}).
The diagram is automatically converted into a Python code that can run in the programmatic framework shown in \figref{fig:code} by the frontend side and sent to a backend server.

The implementation of this frontend interface pursued easy customization and refinement of search queries.
Parameters in queries can be easily tweaked on the blocks.
In addition, they can apply the same queries to a different person in the same video, as well as to other videos.
Then, search results can be compared in line (\figref{fig:interface}D) while jumping between detected scenes.
Also, users can copy and paste search queries by clicking the share button (\figref{fig:interface}G), which exports the internal JSON representation of the diagram.
Here, the JSON representation is merely used for the purpose of sharing queries in the form of diagrams among the users of \proposed{}, and we expect users without programming experiences to neither analyze nor manipulate it directly.

Furthermore, we implemented the inspection page (\figref{fig:interface}E) to foster users' transparent understanding of the search results.
On this page, users can examine how each feature has contributed to the search results and infer which part of the query they need to modify when they find some gaps between their desired results and the actual results.
More specifically, this page presents the results at the level of each block in a manner that shares the same timeline.
Also, users can zoom in and out of the result charts so that they can inspect the work of each block concerning a specific scene.

In addition, this page visualizes how the results would be influenced by changing each parameter, as presented in \figref{fig:interface}F.
Specifically, it shows the relationship between the parameter, the number of detected scenes, and their total duration.
We prepared the chart to help users understand the complicated relationship and adjust the parameters.
For example, it is clear that setting a too-large value makes no scene included in the results.
On the other hand, a too-small value makes \proposed{} detect the entire video as a single consecutive scene.
This nonlinear situation demands the visualization feature to maintain high transparency and customizability.
Specifically, without this feature, users would have difficulties finding the optimal value to increase or decrease the number of scenes to be included in the results.

Lastly, we implemented a report button that is located at the upper right of the video preview on the inspection page.
Using this button, users can remove results that they judge meaningless or mistakenly detected due to inaccurate machine-learning algorithms to obtain features.
The information is also logged in the server.
We believe that such a mechanism allowing continuous improvement of the algorithms in a collective manner is important considering the nature of machine-learning algorithms that cannot assume infallibleness~\cite{DBLP:conf/chi/Yakura23}.

\subsubsection{Backend Server}

Queries composed in the frontend interface are executed in the backend server.
Our implementation puts special emphasis on its processing performance because we expected that it is crucial for enabling iterative customization.
We implemented two background processes dedicated to video processing and query execution.
Once the video processing is completed, all features extracted are cached for iterative query execution.
For example, with the GeForce RTX 3060, we confirmed that the video processing takes approximately 10 minutes for a 1-hour full HD video of 10~FPS (see \secref{app:implementation} for details).
Then, the query execution is completed within approximately 1--2 seconds, which would be sufficient for iterative exploration.
While the query execution time is linearly scalable for the duration of videos and the number of features in the query, given the time taken for video processing, we expect that using it with videos longer than 3--4 hours would be impractical.

\section{User Study}

To evaluate the efficacy of the developed tool in multimodal scene search of conversations, we conducted a within-participant remote user study to compare two conditions: \textit{proposed} and \textit{control}.
Participants used \proposed{} to search scenes within conversational videos in the \textit{proposed} condition.
In the \textit{control} condition, they used a normal video player on their laptop, which was identified as the current standard method in the formative study.
While there could be other baselines, such as letting participants write codes to extract scenes, the above condition is most suitable as we targeted users who do not necessarily have a machine-learning experience.

\subsection{Task}

We first chose two conversational videos from our in-house database in which two people talk via Zoom.
Specifically, we used videos of mock sales and coaching sessions to replicate the situations we were informed about in the formative study.
Their lengths are roughly 12 minutes.
For each participant, we randomly assigned one video to the \textit{proposed} condition and the other to the \textit{control} condition.
Their task was to search for two kinds of scenes from the video.
Here, we prepared four kinds of scenes in total that had been regarded as important scenes in previous literature and can be linked with the practice of reviewing conversations, according to the formative study.
Specifically, we employed scenes, ``when a person looks away and their speech contains filler'' (this can be associated with the intention of taking a turn~\cite{DBLP:conf/iva/PfeiferB09}), ``when a person's speech speed is high and their speech volume is large'' (this can be associated with the expression of fear and anxiety~\cite{Siegman1993}), ``when a person is questioned and it took time in response'' (this can be associated with multiple factors, like planning or hesitation~\cite{Clark2006Pauses}), and ``when a person's speech contains filler and their speech speed is low'' (this can be associated with the repair of not only linguistic errors but also appropriateness errors in conversations~\cite{bosker2013makes}).
The assignment of the target scenes was also randomized and counter-balanced.

\subsection{Participants}

Given our goal of supporting experts in multimodal scene search of conversations, we recruited seven participants who share a similar situation to the participants of our formative study, \ie those who had experiences analyzing conversational videos in their business (\eg giving feedback on recorded sales conversations).
Furthermore, the formative study suggested educational factors in the motivation of reviewing conversation videos; some wanted to transfer their knowledge and perspective of reviewing to junior or other members.
Thus, we assumed that it would also be plausible that \proposed{} can support novice people.
Therefore, we also recruited five participants who had no experience conducting scene search of conversations and observed the difference in the efficacy of \proposed{} for experts and novices.
In total, we recruited 12 participants (P1 -- P12; 9 male, 3 female; aged between 24 and 43, $M = 30.0$, $SD = 6.4$).
We note that nine of them had no or less than half-a-year experience in machine learning.

\subsection{Procedure}

After completing a demographic questionnaire, the participants accessed the web-based tool and were instructed on how to use the tool with a 2-minute tutorial video.
Following the video, they were guided to make one query that was not relevant to the task scenes as a practice.
This practice took approximately seven minutes, in which the participants got familiar with \proposed{}'s functions.
Then, they worked on the task under either \textit{proposed} or \textit{control} conditions.
The order was counter-balanced.

In the \textit{control} condition, they watched a video without the tool.
Specifically, we asked them to use QuickTime Player\footnote{\url{https://support.apple.com/downloads/quicktime}} to watch as usual. 
They were also asked to write down the start and end times of each scene they thought matched the given descriptions.
In the \textit{proposed} condition, they composed blocks to find the scenes without necessarily watching the whole video.
They were asked to tune the queries until they thought the result scenes matched the given descriptions.
Since participants were required to comprehend what features were available in \proposed{}, we prepared a document listing them so that the participants could refer to it anytime.
We also had the video preprocessed in \proposed{} so that they would not need to wait for the inference of the video processing module.

After the task, they filled out a questionnaire about cognitive load.
In the \textit{proposed} condition, they also answered a questionnaire about usability, and in addition, they watched the entire video manually to examine whether the extracted scenes actually matched the given descriptions as well as whether there were any missing scenes, answering their satisfaction.
We also asked for their open-ended comments at the end.
They repeated the task under the other condition with a new video and scene descriptions to search for.
It took roughly 90 minutes for a participant to finish the study.

\subsection{Results}

\subsubsection{Usability}

Participants answered the System Usability Scale (SUS)~\cite{brooke1996sus} in the \textit{proposed} condition.
The mean score for SUS was $74.4$~(SD $= 17.2$), which indicated that the tool had reasonably good usability for users.
The score was higher for the novice users (Mean $= 77.0$, SD $= 18.1$) than the experienced users (Mean $= 72.5$, SD $= 17.7$), implying that the computational support of \proposed{} was also effective for people who do not necessarily have knowledge or experience, while we did not see statistically significant differences in the scores.
Participants commented positively on the user experience of the tool, especially the design of the visual programming interface: \textit{``This tool allowed me to instantly see the result of queries and try refining my query iteratively''} [P9], and \textit{``I enjoyed doing the task as if it had been a puzzle''} [P6].
At the same time, adjusting the threshold to select scenes by features was observed to be the most complex process in the use of \proposed{}.
Here, some participants mentioned that the visualization function was helpful: \textit{``The detail page (\figref{fig:interface}E) gave me a clear understanding of the filtering logic''} [P11].
We observed that this participant first ran the search with a rough set of thresholds and then visited the inspection page to refine the thresholds.
\textit{``I at first struggled to grasp how the threshold changes the number of scenes to be detected. The graph (\figref{fig:interface}F) was helpful''} [P12].
These comments collaborate with our interpretable design of \proposed{}.

\subsubsection{Cognitive load}

\begin{figure*}[t]
    \begin{center}
        \includegraphics[width=\textwidth]{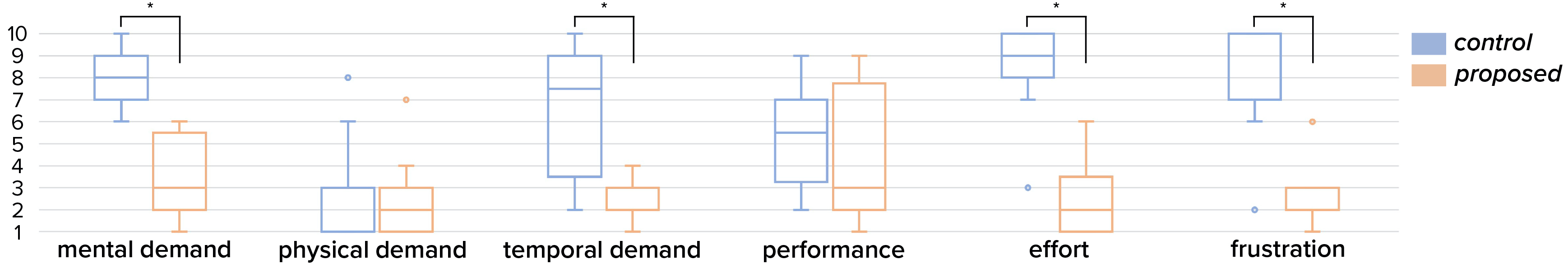}
    \end{center}
    \caption{Participants' evaluation of their cognitive load in the user study. $p < .05$ is marked as *.}
    \label{fig:user_study_nasatlx}
\end{figure*}

We also measured participants' cognitive load using six factors from the NASA-TLX~\cite{hart1988development}, which is shown in~\figref{fig:user_study_nasatlx}.
Results of paired t-test indicated that participants felt less mental demand ($t(22) = 8.232,~p = 4.971 \cdot 10^{-6},~\text{Cohen's }d = 2.376$), temporal demand ($t(22) = 5.745,~p = 0.00013,~\text{Cohen's }d = 1.658$), effort ($t(22) = 6.636,~p = 3.678 \cdot 10^{-5},~\text{Cohen's }d = 1.916$), and frustration ($t(22) = 5.402,~p = 0.00022,~\text{Cohen's }d = 1.559$) in the \textit{proposed} condition than in the \textit{control} condition.
We observed the same tendency when we analyzed the results separately for novice and experienced participants; both exhibited significant reductions in mental demand, temporal demand, effort, and frustration regarding the \textit{proposed} condition.
Participants mentioned difficulty performing the task in the \textit{control} condition: \textit{``I was struggling to maintain concentration, and I had less satisfaction because I was concerned that I missed detailed signs''} [P3].
Especially, it was difficult for them to pay attention to differences in quantitative values of multimodal signals: \textit{``It was hard to identify the differences in volume and speech speed''} [P1].
On the other hand, participants found \proposed{} efficient: \textit{``The good point about the interface was that I could efficiently search scenes from videos.''} [P7], and \textit{``I had much less workload because I could choose scenes from automatically extracted ones by the algorithm''} [P9].

\subsubsection{Task completion time}

\begin{table}[t]
  \caption{Comparison of the task completion time (in seconds).}
  \label{tbl:user_study_time}
  \begin{tabular}{cccc}
    \toprule
    Condition & All participants & Experienced & Novice \\
    \midrule
    \textit{control} & $1021.58$ ($\pm 409.57$) & $993.29$ $(\pm 471.97)$ & $1061.20$ $(\pm 407.11)$ \\
    \textit{proposed} & $766.75$ ($\pm 327.47$) & $773.14$ ($\pm 333.53$) & $757.80$ $(\pm 393.24)$ \\
    \bottomrule
  \end{tabular}
\end{table}

We compared task completion time between the \textit{control} and \textit{proposed} conditions.
Although the mean time in the \textit{proposed} condition was lower than in the \textit{control} condition, the difference was not statistically significant ($t(22) = 1.704,~p = 0.0582$), which was consistent also when we separately examined the completion time of the experienced and novice participants.
When we inspected individual results, two novice participants took much time to find appropriate features for the tasks in the \textit{proposed} condition ($1152.0$ and $1418.0~[s]$, respectively), which is also inferred from their comment, \eg \textit{``I thought I needed to understand prerequisite knowledge about features''} [P5].
This suggests room for improvement, that is, adding a brief explanation of each feature on the interface so that users can check it instantly.

Four participants tried to find the given two kinds of scenes separately in the \textit{control} condition, while others did it simultaneously.
When we only look at the completion time of those four participants, $t$-test shows a significant difference ($t(6) = 6.241,~p = 0.0041,~\text{Cohen's }d = 3.120$) between the \textit{control} ($M = 1545.50$, $SD = 127.58$) and \textit{proposed} conditions ($M = 740.00$, $SD = 322.18$).
We conjecture that some participants could find scenes corresponding to multiple descriptions of target human behaviors simultaneously, while they needed to make queries sequentially in the \textit{proposed} condition.
On the other hand, this is only applicable when the number of descriptions is limited, while the computational support of \proposed{} is scalable.
This result implies that \proposed{} can reduce the time to find scenes corresponding to a single description.

\subsubsection{Satisfaction}
\label{sec:study-results-satisfaction}

\begin{table}[t]
  \caption{Participants' satisfaction on outputs of the \proposed{} in the \textit{proposed} condition. (5-point Likert scale)}
  \label{tbl:user_study_satisfaction}
  \begin{tabular}{lccc}
    \toprule
    Question & All participants & Experienced & Novice \\
    \midrule
    \makecell[l]{I could extract enough scenes\\I wanted to find}
    & $4.00$ ($\pm 1.21$) & $4.29$ ($\pm 0.76$) & $3.75$ ($\pm 1.89$) \\
    \midrule
    \makecell[l]{Scenes extracted by the tool\\were not meaningful}
    & $2.17$ ($\pm 1.27$) & $1.86$ ($\pm 0.69$) & $2.75$ ($\pm 2.06$) \\
    \bottomrule
  \end{tabular}
\end{table}

Participants answered two questions about their satisfaction with the outputs: ``I could extract enough scenes I wanted to find'' and ``Scenes extracted by the tool were not meaningful'' with a 5-point Likert scale.
These questions measure satisfaction in terms of recall or precision, respectively.
\tabref{tbl:user_study_satisfaction} shows the results.
They felt the extracted scenes covered enough scenes, commenting \textit{``It served as an objective basis for analysis, as it performs searches based on clear parameter thresholds, compared to seeing videos with my subjective judgments''} [P5].
It was notable that the experienced participants specifically appreciated the support of \proposed{}, which can be attributed to the challenges they have faced.
While they found the results meaningful, a few participants mentioned room for improving some algorithms calculating features, \textit{``I felt a few extra scenes were included, maybe due to an error in speaker identification''} [P9].
Since \proposed{} is modular regarding its feature-extraction algorithms, further refinement (in this case, speaker diarization) will help resolve the issues.

\subsection{Discussion}

Overall, participants felt positive about \proposed{}'s usability, experienced a less cognitive load, and were satisfied with the results extracted by the tool.
Furthermore, the efficacy was confirmed not only for the experienced participants but also for the novice participants who do not necessarily have knowledge for conducting multimodal scene search.
We found that this was largely attributed to the design considerations we extracted from our formative study, particularly \textit{customizability} and \textit{transparency}.

Customizability seemed to contribute to efficiently refining queries toward enumerating desired scenes from the videos through a nonlinear, trial-and-error process.
Interestingly, such an iterative and explorative process stimulated participants to imagine the tool's applicability outside the experimental setting; P6 mentioned that he would like to use \proposed{} for monitoring speech characteristics, such as the frequent phrases they use.
We want to emphasize that this would not be observed unless we offered various features, as listed in \tabref{tbl:features}, with an intuitive interface for leveraging them.
At the same time, this wide range of available features demanded the participants familiarize themselves with the tool, which affected its usability and the completion time.
We believe that it can be eased by adding explanations on the interface as well as providing the users with examples, as Ichinco~\etal~\cite{DBLP:journals/jvlc/IchincoHK17} suggested the importance of example code in helping novice users in visual programming systems.

Transparency seemed to affect the cognitive load and the participants' satisfaction with the results.
Particularly, the design of \proposed{} that uses explicit parameters rather than black-box approaches serves as an objective foundation of the analyses, corroborating prior findings~\cite{DBLP:conf/chi/WuTC22, RapsaiDu}.
Notably, this guided the participants to minutely interpret the results, as P9 commented on the probable cause for false positives (see \secref{sec:study-results-satisfaction}).
Such interpretability would not be available without clear visualization of how multimodal signals were combined, and it would allow users to fill in discrepancies between their intentions and results iteratively.

\subsubsection{Limitation}

We also note that the design of this study has several limitations.
First, the \textit{control} condition was intended to replicate the usual workflow of the participants, and we did not strictly control how to perform it.
Still, due to the study environment, we asked them to use QuickTime Player, which features common video playback functions such as adjusting speed.
There can be a chance that some participants were not very familiar with it, \eg not knowing shortcuts, which might have affected the results. 
Furthermore, while the context of the videos used in this study was chosen to be linked to the use cases mentioned in the formative study, the videos and tasks did not cover all of the possible use cases.
This point partially motivated us to observe the in-the-wild use of \proposed{}, which is reported in the following section.

\section{In-the-Wild Use of \proposed{}}
\label{sec:wild}

While our user study exhibited the benefits of its customizability and transparency, \proposed{} is also designed to promote query reusability, which allows various queries composed by users to be shared as knowledge of behavior analysis.
This has a large potential to help experts' analysis work, which currently depends on their individual skills, as we discovered in \secref{sec:formative}.

To examine how the tool affects their workflow, we invited 11 experts to its trial through our personal contacts from industrial organizations that regularly conduct human behavior analysis.
There were six coaches in executive coaching, two assessors in human assessment, and three salespeople.
Their background and expertise are summarized in \tabref{tbl:interviewees-deployment}.

\begin{table}[t]
  \caption{Backgrounds of the experts in the in-the-wild trial. \textbf{Co}, \textbf{As}, and \textbf{Sa} stand for coach, assessor, and salesperson, respectively.}
  \label{tbl:interviewees-deployment}
  \begin{tabular}{lll}
    \toprule
    ID & Gender & Job experience \\
    \midrule
    Co1  & M & 6 years \\
    Co2  & M & 6 years \\
    Co3  & F & 10+ years \\
    Co4  & M & 10+ years \\
    Co5  & F & 2 years \\
    Co6  & F & 2 years \\
    As1  & M & 2 years \\
    As2  & M & 5 years \\
    Sa1  & M & 3 years \\ 
    Sa2  & F & 3 years \\
    Sa3  & M & 7 years \\  
    \bottomrule
  \end{tabular}
\end{table}

We deployed \proposed{} as a Web application, and the participants could access it during the trial period.
Specifically, they could upload their own conversational videos and use \proposed{} for analysis. 
We stored neither the raw data nor the extracted features in an accessible format since it contains privacy-sensitive data (\eg meeting with clients).
The study length was 3 weeks (the precise period varied slightly for each participant based on their work schedule).
The participants could ask experimenters for questions via text communication if needed.
After the trial, we conducted unstructured interviews with each participant with regard to how the tool affected their workflow and what kind of additional features they wanted.
We obtained approval from our institution to conduct the study.

\subsection{Knowledge-Share Repository}
\label{sec:wild-repository}

\begin{figure*}[t]
    \begin{center}
        \includegraphics[width=\textwidth]{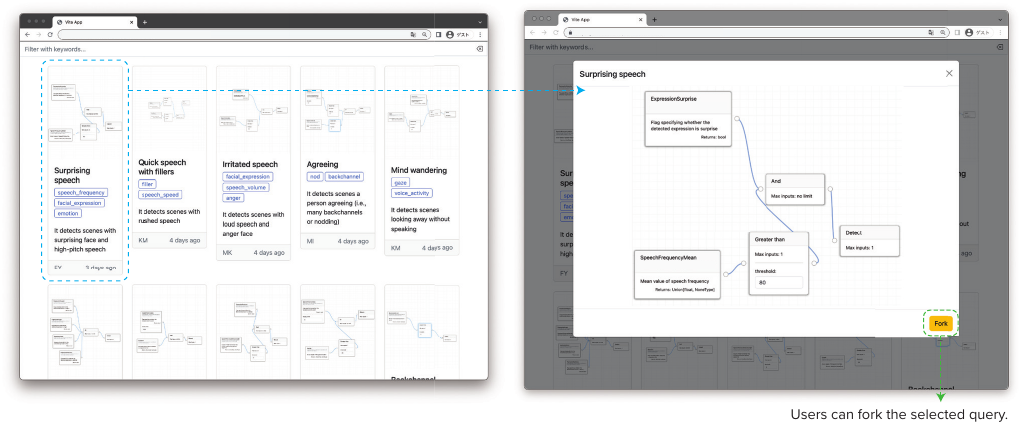}
    \end{center}
    \caption{Knowledge-share repository of \proposed{} where users can contribute and access queries.}
    \label{fig:repository}
\end{figure*}

To facilitate the reuse of the queries, we developed a knowledge-share repository where users can contribute their queries and refer to ones shared by others.
The repository interface is shown in \figref{fig:repository}.
Users can directly export their created queries to this repository with a single button (\figref{fig:interface}G).
Each query is associated with its semantic meaning describing behavioral cues, such as ``get stuck in answering questions,'' and users can look for queries in the repository by text description or used features (\figref{fig:repository} left).
They can fork the selected query, edit it freely, and execute it on their end (\figref{fig:repository} right).
This repository was designed to serve as an example code base to help novice users familiarize themselves with the tool.
We initially populated the repository with several sample queries as references.
Then, we hosted the repository at different Web URLs for each organization of the participants by preparing different databases.
In other words, their created queries were only shared among those from the same organization.

\subsection{Results}
\label{sec:wild-results}

\subsubsection{Generated Queries}

After all trials, we manually examined the queries and their described meaning in the knowledge-share repository\footnotemark.
\footnotetext{We did not access the raw video data, only checking the queries.}
Several queries were duplicated regarding the feature relationship; after removing the duplication, we had 22 semantically-different queries.
Additionally, we had an external expert in human behavior analysis validate the queries and their semantic meanings, which confirmed the queries generated by the participants were reasonable in terms of the feature relationship.
These results reaffirmed the diverse scenarios of behavior analysis and the effectiveness of \proposed{} in generating and sharing queries.
For instance, some queries involved eight blocks to combine three different behavioral modalities, which would be difficult to explore and customize in ways other than visual programming, such as parametric design, which would result in selecting and tweaking appropriate ones from all available parameters.
Some examples of the queries shared through this process are presented in \figref{fig:example-queries} as an appendix.

\subsubsection{How \proposed{} Affected the Workflow of Experts}

In their comments, the participants first agreed that \proposed{} allows them to save time.
As1 mentioned, \textit{``I use the tool whenever I need to watch the videos to analyze clients' behavior during the assessment session. I like it because I don't have to watch the entire video anymore, especially when I have conducted the session and know what was talked about. I usually apply two or three queries and watch the extracted scenes to make a decision.''}
Similarly, Sa2 said, \textit{``I use the tool to analyze more about clients' behaviors after reviewing the transcription of the sales meetings first. The scene extraction mechanism matches with my intuition. In watching the sales videos, I want to focus on the other person's reactions. For example, just watching the moments when it took longer for them to answer is informative. This process was originally very tedious and I often got distracted while watching videos.''}
From these comments, we confirmed the advantage of introducing computational support in this burgeoning domain by aiding its time-consuming, manual process.

Moreover, the participants valued the objectivity the tool brings to their workflow.
As2 mentioned, \textit{``The tool was intuitive. Recently, communication within the organization has been changing, I would say, from being subjective to objective. Assessors try to verbalize their own knowledge using behavior features and sometimes use the extracted scenes as evidence.''} 
These participants started incorporating the tool into their workflow, leveraging it to facilitate the analysis or create a foundation for objective communication between experts.
We also found that there were several queries that experts in the same organization often apply as a routine starting point.
For example, salespeople use a query to see moments when it takes longer to answer questions, as Sa2 commented.
Sa1 commented on the benefit of using consistent criteria to assess multiple sales conversations, \textit{``We use the same queries across multiple videos and compare the results. It is helpful to understand the differences between good and bad meetings quantitatively.''}
As1 mentioned an idea for a potential interface that could further help their work, \textit{``It would be beneficial to have a video player that allows users to register frequently used queries so that detected sections can be easily referenced while watching the video, similar to YouTube's video chapters.''}
It was suggested that they identified knowledge commonly agreed upon as insightful within processes that were previously reliant on individual manual labor, and they have streamlined these through \proposed{}.

In addition, Co2 mentioned potential ways for educating junior coaches, saying, \textit{``If you're a new coach, your mentor checks the session video. That process can be facilitated by extracting pre-defined behaviors with this tool in such a transparent manner. I find it fascinating that we can turn tacit knowledge into formal knowledge regarding how to evaluate coaching skills.''} 
Sa3 mentioned a potential system for developing their communication skills, \textit{``We are thinking of creating a real-time feedback system that alerts us when we behave badly as we define, like not looking at the front camera while speaking.''}
As we described in \secref{sec:proposed-impl-frontend}, \proposed{} supports easy query export as a standalone detection program, so it is possible to realize these educational applications.

Additionally, participants mentioned that they enjoyed building queries through the visual programming interface while referring to and being inspired by the queries already shared in the knowledge-share repository.
Given that the participants generated semantically diverse queries, we suppose \proposed{} contributed to their creativity in uncovering insights into human behaviors by democratizing the method for the analysis with a low-code approach, as discussed in ~\cite{OreillyLowCode}.

\subsubsection{Room for Further Improvements}
\label{sec:wild-results-room}

The participants mentioned that the knowledge-share repository and the visualization of the raw features were helpful for them to use \proposed{}.
In detail, the repository helped them create the first few queries by serving as references, and the visualization facilitated their tuning of the queries.
On the other hand, some reported difficulty associating the high-level concept of human behavior with low-level representations, \ie a query in \proposed{}.
Co5 commented, \textit{``I sometimes found scenes with features that were difficult to verbalize. In such cases, it was hard to come up with the feature combination.''}
This underpins the general challenge in human behavior analysis; the knowledge is not well structured and is highly dependent on each expert implicitly~\cite{DBLP:journals/corr/abs-2204-08471}.
We discuss potential approaches to addressing this issue in \secref{sec:discussion-semiautomating}.

\subsection{Discussion}

The result that \proposed{} introduced objectivity to the traditionally subjective domain represents a pivotal advantage of computer-supported tools, which is evidenced across various sectors of expert-AI interactions in prior work, such as satellite imagery analysis~\cite{DBLP:journals/pacmhci/MorrisonSHP23} and clinical diagnostic support~\cite{DBLP:journals/pacmhci/GuHHC21, DBLP:journals/asc/WangJYXLWZHMXZS21, DBLP:journals/imwut/ArakawaAMTSLG23}.
For example, Morrison~\etal~\cite{DBLP:journals/pacmhci/MorrisonSHP23} identified multiple strategies for explainable AI to support assessors in building damage assessment from satellite imagery, such as by presenting causal explanations for its inference or by showing the number of damaged buildings in an image to indicate the severity.
As another example, Impetus~\cite{DBLP:journals/pacmhci/GuHHC21} is designed to incrementally support pathologists as they analyze whole-slide images to detect tumors.
In its initial phase, the tool highlights potential areas of concern via anomaly detection.
Subsequently, these delineated areas are magnified, allowing pathologists to make informed decisions based on AI inferences.
The principle of such phased mixed-initiative~\cite{DBLP:conf/chi/Horvitz99} was also contemplated during the development of \proposed{}.

Here, distinctions arose due to 1) an open-ended nature of the tasks with the limited structured knowledge in the burgeoning domain of conversational analysis and 2) the domain's inherent dependency on subjective judgment, requiring special attention for automating~\cite{DBLP:journals/corr/abs-2204-08471, DBLP:conf/chi/ParkAHL22}.
These points make it difficult to define a clear goal that is tractable for AIs, and thus, we did not aim to develop individual machine-learning models to estimate final outcomes (\eg assessing a candidate's level) or to replace specific parts of the tasks.
Rather, we emphasize exploration by extracting key conversational moments and facilitating sharing of these insights on human behavior.
We intended to help experts translate their subjective judgment into consistent criteria and extract structured and portable knowledge.
Our in-the-wild trial demonstrated varied utilization of the proposed tool within individual organizations, resulting in evolving workflows.
This result implies that the plausible directions of an expert-centered approach are not only limited to immediately solving their tasks with a dedicated machine learning model but include helping their in-process knowledge organization and its sharing.

\section{Limitations and Future Work}

Through the two studies, we have confirmed the effectiveness of \proposed{}'s design in multimodal scene search and demonstrated how such a tool transforms experts' workflow in the conversational analysis domain.
Lastly, we discuss future work that can enhance \proposed{}'s effectiveness and application possibilities.

\subsection{Calibrating Features}

Reusing shared queries may necessitate calibration of certain features concerning their detection thresholds.
For instance, gaze direction depends on camera placement, and a threshold determined based on other videos might not be directly applicable.
Thus, we plan to indicate which features require calibration on the diagram.
We anticipate that the time-series visualization of raw features (\figref{fig:interface}E, F) will assist users in adjusting threshold values.

\subsection{Expanding Features}
\label{sec:discussion-features}

While our initial selection covered a diverse set of multimodal features by accounting for the needs found in the formative study, we plan to expand it.
In the comments of the in-the-wild trial, we observed the participants requesting specific features, such as detecting if a hand is on a chin because this gesture implies boredom~\cite{pease1981body}.
The modularity of \proposed{} enables us to integrate such new algorithms as feature blocks.
Moreover, the current tool does not consider time-series order relationships, such as the transition of a person's facial expression from happy to angry.
Some participants expressed interest in adding such order constraints to queries to enhance search capability.
We believe these suggestions are the fruit of the participants' interaction with \proposed{} that enabled them to imagine the possibility of utilizing machine learning algorithms.
Given that, maintaining the development of the tool with their requests will contribute to the behavior analysis research in a collective manner.

\subsection{Analyzing Group Dynamics of Conversation}

Our research focuses on individual behaviors that are essential in the domain of conversational analysis, which was suggested in the formative study.
However, the interplay among multiple people (\eg a situation where a person gets uncomfortable due to what his conversation partner said) is not well-considered in the current implementation.
We can address these group dynamics by changing data structures and adding a logic operator that handles time-series order constraints, as mentioned in \secref{sec:discussion-features}.
This logic operator will allow users to analyze conversational videos based on multiple group members' multimodal behaviors, such as predicting influential statements or functional roles in group discussions \cite{DBLP:conf/icmi/DongLCPPZ07, DBLP:conf/icmi/NiheiNHHO14}.

\subsection{Semi-Automating Generation of Queries}
\label{sec:discussion-semiautomating}

We also mention that the difficulty of behavior analysis is attributed to its knowledge (\ie translating scene semantics into low-level feature representation) being often implicit and subjective, as discussed in~\secref{sec:wild-results-room}.
Therefore, we endeavor to make the process more easy, transparent, and programmable in future work.
Here, the knowledge accumulated as reusable queries can further support the analysis process, \eg semi-automating the query generation.
The tool can list relevant features given the semantic descriptions by leveraging the queries accumulated in the knowledge-share repository.
Moreover, we can model the process of generating queries that will result in extracting sample scenes specified by users as an inverse problem \cite{Tarantola2005}.
We are eager to investigate these directions and to understand users' mental models during the analysis process with \proposed{}, which will construct better forms of expert-AI collaboration in behavior analysis.

\subsection{Clarifying Potential Influence}
\label{sec:discussion-potential}

It is an important future work to investigate the potential influence of \proposed{} on the workflow of a broader range of human behavior analysis.
While our current in-the-wild trial examines it with experts from coaching, assessment, and sales communication, there can be more diverse conversation contexts where experts share the need identified in \secref{sec:formative-conclusion} and \proposed{} can be applied.
Simultaneously, each use case should be carefully designed due to possible adverse effects.
For instance, as a machine-learning-based system, there can always be the risk of false results; false negatives can prevent some key scenes from being considered by experts.
It is easy to anticipate that fully automating the process with \proposed{} (\eg job interview) is not ideal, as pointed out by Arakawa and Yakura~\cite{DBLP:journals/corr/abs-2204-08471}.
Thus, we keep studying how \proposed{} affects workflow in different organizations in the longer term.

\section{Conclusion}

We presented \textit{\proposed{}}, a multimodal scene search tool for conversations, which enables users to generate various queries by combining diverse features (\ie non-verbal, para-linguistic, and linguistic) of human behavior easily through a visual programming interface.
Our semi-structured interviews with experts in diverse scenarios confirmed a common need for computational support in analyzing human behavior in conversational videos and highlighted key design considerations, which informed \proposed{}'s development.
The user study demonstrated its effectiveness in reducing cognitive load, exhibiting favorable usability with satisfaction in the outputs, and showing a trend towards decreasing search completion time.
Moreover, by deploying the tool in actual scenes of behavior analysis, we found that the objectivity it brings can affect experts' workflow, suggesting further useful applications in multiple contexts.
We believe our work paves the way for establishing expert-AI collaboration in a burgeoning domain, supporting their analysis and exploration that require a significant amount of human context through a dedicated tool.

\begin{acks}
This work was supported in part by JST ACT-X Grant Number JPMJAX200R and JSPS KAKENHI Grant Numbers JP21J20353.
\end{acks}

\bibliographystyle{ACM-Reference-Format}
\bibliography{paper}

\appendix

\newpage
\section{Appendix}

\subsection{Implementation Details for Behavior Features in Providence}
\label{app:implementation}

\tabref{tbl:appendix_features} shows the implementation details and inference speed of each machine-learning model we used in \proposed{}.
We also noted the accuracy information on a public or our in-house dataset if available.

\begin{table}[H]
\caption{Details of the implementation for multimodal features in \proposed{}.}
\label{tbl:appendix_features}
\scriptsize
\renewcommand{\arraystretch}{1.1}
\begin{tabularx}{\linewidth}{lXl}
\toprule
Feature               & Details & Inference speed \\
\midrule
Facial keypoints      & Facial keypoints are extracted by the RMPE model~\cite{DBLP:conf/iccv/FangXTL17}. Our model was trained with the 300W dataset~\cite{DBLP:conf/iccvw/SagonasTZP13}. Its normalized mean error on the dataset is \SI{5.53}{\percent}. & \SI{25.9}{\milli\second} per frame \\
Body pose             & Body pose is estimated by the AlphaPose model~\cite{alphapose}. Our model was trained with the COCO-Pose dataset~\cite{DBLP:conf/eccv/LinMBHPRDZ14}. Its mean average precision on the dataset is \SI{72.2}{\percent}. & \SI{52.5}{\milli\second} per frame \\
Head pose             & Head pose is estimated by solving the perspective-n-point problem using facial keypoints. & \SI{27.5}{\milli\second} per frame \\
Gaze direction        & Gaze direction is estimated by the RT-GENE model~\cite{DBLP:conf/eccv/FischerCD18}. Our model was trained with the dataset introduced in~\cite{DBLP:conf/eccv/FischerCD18}. Its mean angle error on the dataset is \SI{10.4}{\degree}. & \SI{33.6}{\milli\second} per frame \\
Facial expression     & Facial expression is classified by the ResNetAttention model \cite{DBLP:conf/cvpr/HeZRS16} into seven categories: anger, disgust, fear, happiness, sadness, surprise, and neutral. Our model was trained with an in-house facial expression dataset. Its accuracy on the dataset is \SI{69.1}{\percent}. & \SI{16.9}{\milli\second} per frame \\
Nod                   & Nod is detected with a rule-based algorithm based on how rapidly facial keypoints are moving in a certain time window. & \SI{3.80E-1}{\milli\second} per frame \\
Blink                 & Blink is detected with the EAR (eyes aspect ratio) algorithm using facial keypoints. Its F1 score on an in-house dataset is \SI{78.0}{\percent} & \SI{2.08}{\milli\second} per frame \\
Speech text           & Speech text is transcribed from audio using the Conformer model~\cite{DBLP:conf/interspeech/GulatiQCPZYHWZW20}. Our model was trained with LaboroTV dataset \cite{DBLP:conf/icassp/AndoF21} and in-house conversational dataset. Its CER (Character Error Rate) on the datasets is \SI{10.5}{\percent}. & 0.19 (real-time factor) \\
Filler \& backchannel & Fillers and backchannels are detected with a pattern-matching algorithm by comparing transcriptions with a pre-defined list of filler and backchannel words. & \SI{3.66E-3}{\milli\second} per sentence \\
Voice activity        & Voice activity is detected using \textit{pyannote-audio}\footnotemark{}. Our model was trained with four datasets: AMI\footnotemark{}, AISHELL4\footnotemark{}, VOXCONVERCE\footnotemark{}, and RIRS\_NOISES\footnotemark{}. & \num{7.8E-2} (real-time factor) \\
Question              & Sentences are judged whether they contain a question or not by the \textit{AutoModelForSequenceClassification} model, one of the classification models in huggingface\footnotemark{}, and the BERT tokenizer~\cite{DBLP:conf/naacl/DevlinCLT19}. Both the classification model and the tokenizer were trained with Tanaka Corpus\footnotemark{}. Its accuracy on an in-house dataset is \SI{90.7}{\percent} & \SI{6.40}{\milli\second} per sentence \\
Sentiment             & Sentences are also classified into positive, negative, or neutral with a BERT-based classification model ~\cite{DBLP:conf/naacl/DevlinCLT19}. This model was trained with a subset of multilingual-sentiments\footnotemark{} datasets. Its accuracy on the datasets is \SI{83.3}{\percent} & \SI{6.24}{\milli\second} per sentence \\
Speech frequency      & Fundamental frequency, \textit{F0}, is calculated from speech wave. & \num{4.6E-3} (real-time factor) \\
Speech volume         & Speech volume is calculated as the average amplitude of the speech wave in a certain time window. & \num{1.3E-4} (real-time factor) \\
Speech length         & Speech length is calculated as the duration of each audio segment. & \SI{4.06E-2}{\milli\second} per utterance \\
Speech speed          & Speech speed is calculated by dividing the number of characters by the length of a speech segment. & \SI{5.31E-1}{\milli\second} per utterance \\
Speech overlap        & Speech overlaps are detected if speech segments are overlapped among different speakers. The segments are separated based on voice activity for each speaker. & \SI{5.32E-3}{\milli\second} per utterance \\
Speech interval       & Speech interval is calculated by subtracting end time of the previous speech segment from start time of the current speech segment. & \SI{8.37E-4}{\milli\second} per utterance \\
\bottomrule
\end{tabularx}
\end{table}

\addtocounter{footnote}{-7}
\footnotetext{https://github.com/pyannote/pyannote-audio}
\addtocounter{footnote}{+1}
\footnotetext{https://github.com/pyannote/AMI-diarization-setup}
\addtocounter{footnote}{+1}
\footnotetext{https://github.com/FrenchKrab/aishell4-pyannote}
\addtocounter{footnote}{+1}
\footnotetext{https://github.com/joonson/voxconverse}
\addtocounter{footnote}{+1}
\footnotetext{https://www.openslr.org/28/}
\addtocounter{footnote}{+1}
\footnotetext{https://huggingface.co/docs/transformers/model\_doc/auto}
\addtocounter{footnote}{+1}
\footnotetext{http://www.edrdg.org/wiki/index.php/Tanaka\_Corpus}
\addtocounter{footnote}{+1}
\footnotetext{https://huggingface.co/datasets/tyqiangz/multilingual-sentiments}

\newpage
\subsection{Example Queries Generated in the In-The-Wild Trial}
\label{app:workshop}

\begin{figure*}[h]
    \begin{center}
        \includegraphics[width=0.9\textwidth, trim=0 40 0 0, clip]{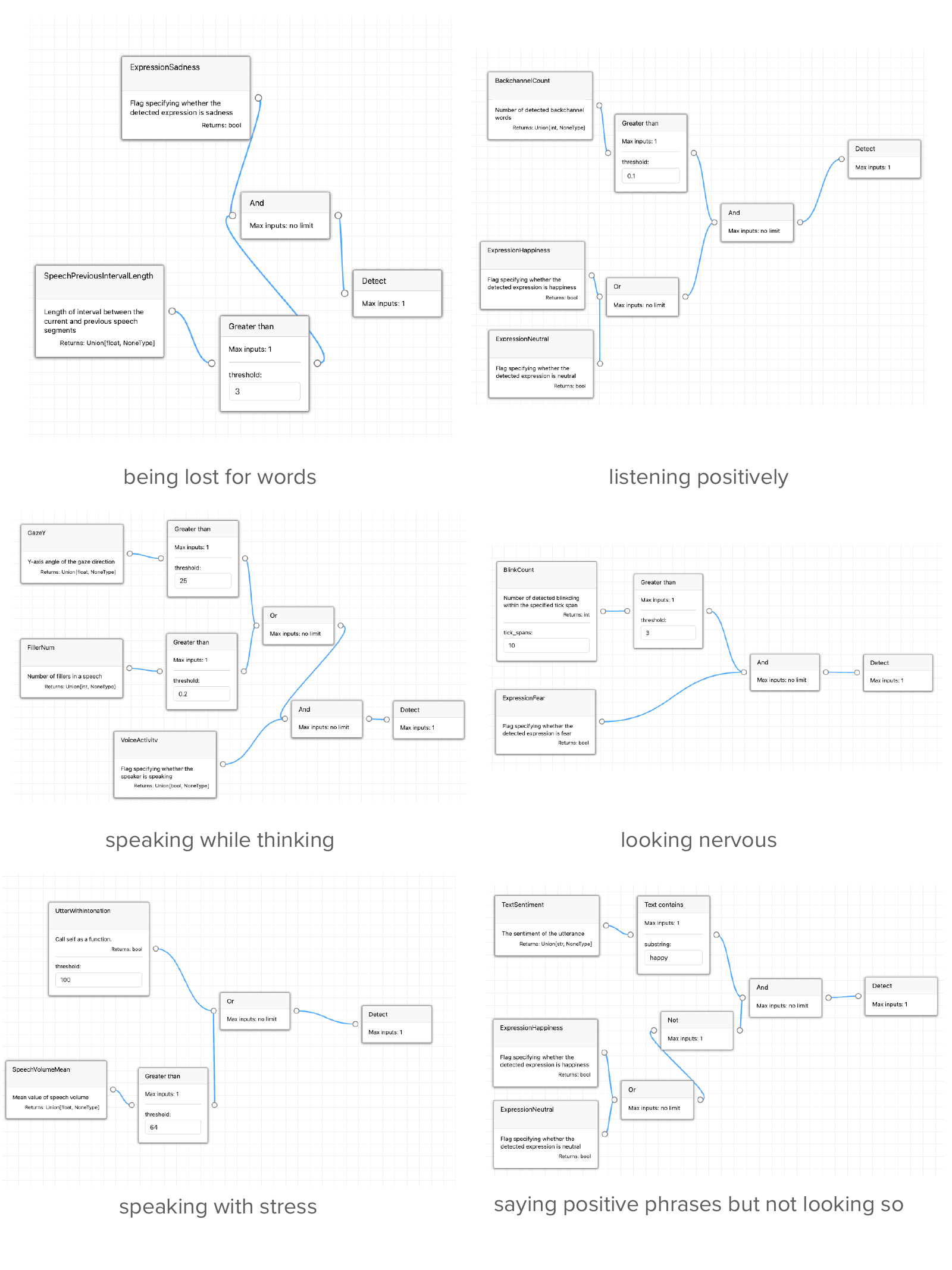}
    \end{center}
    \caption{Example queries participants created, which capture diverse semantics of human behaviors.}
    \label{fig:example-queries}
\end{figure*}

\end{document}